\documentclass[prd,amsmath,amssymb,superscriptaddress,12pt]{revtex4}
\usepackage{graphicx}
\usepackage{dcolumn}
\usepackage{bm}
\newcommand{\beq}{\begin{equation}}
\newcommand{\eeq}{\end{equation}}
\newcommand{\ba}{\begin{array}{ccc}}
\newcommand{\ea}{\end{array}}
\newcommand{\nn}{\nonumber \\}
\newcommand{\br}{{\bm r}}
\newcommand{\bk}{{\bm k}}
\newcommand{\bp}{{\bm p}}
\newcommand{\bq}{{\bm q}}

\newcommand{\bQ}{{\bm Q}}
\def\bea{\begin{eqnarray}}
\def\eea{\end{eqnarray}}
\usepackage{float}
\preprint{}

\begin{document}
\title{Angular fluctuations of a multi-component order\\ describe the pseudogap regime of the cuprate superconductors} 

\author{Lauren E. Hayward}
\affiliation{Department of Physics and Astronomy, University of Waterloo, Ontario, N2L 3G1, Canada}

\author{David G. Hawthorn} 
\affiliation{Department of Physics and Astronomy, University of Waterloo, Ontario, N2L 3G1, Canada}

\author{Roger G. Melko}  
\affiliation{Department of Physics and Astronomy, University of Waterloo, Ontario, N2L 3G1, Canada}
\affiliation{Perimeter Institute for Theoretical Physics, Waterloo, Ontario N2L 2Y5, Canada}

\author{Subir Sachdev}
\affiliation{Department of Physics, Harvard University, Cambridge MA 02138}


\begin{abstract}
The hole-doped cuprate high temperature 
superconductors enter the pseudogap regime as their superconducting critical
temperature, $T_c$, falls with decreasing hole density. Experiments have
probed this regime for over two decades, but we argue that decisive new
information has emerged from recent X-ray scattering experiments
\cite{keimer,chang,hawthorn}. The experiments observe incommensurate
charge density wave fluctuations whose strength rises gradually over
a wide temperature range above $T_c$, but then decreases as the temperature is lowered 
below $T_c$. We propose a theory in which the superconducting and charge-density
wave orders exhibit angular fluctuations in a 6-dimensional space. The theory
provides a natural quantitative fit to the X-ray data, and can be a basis for
understanding other characteristics of the pseudogap.
\end{abstract}

\maketitle

\begin{figure}
\centering
\includegraphics[width=6.5in]{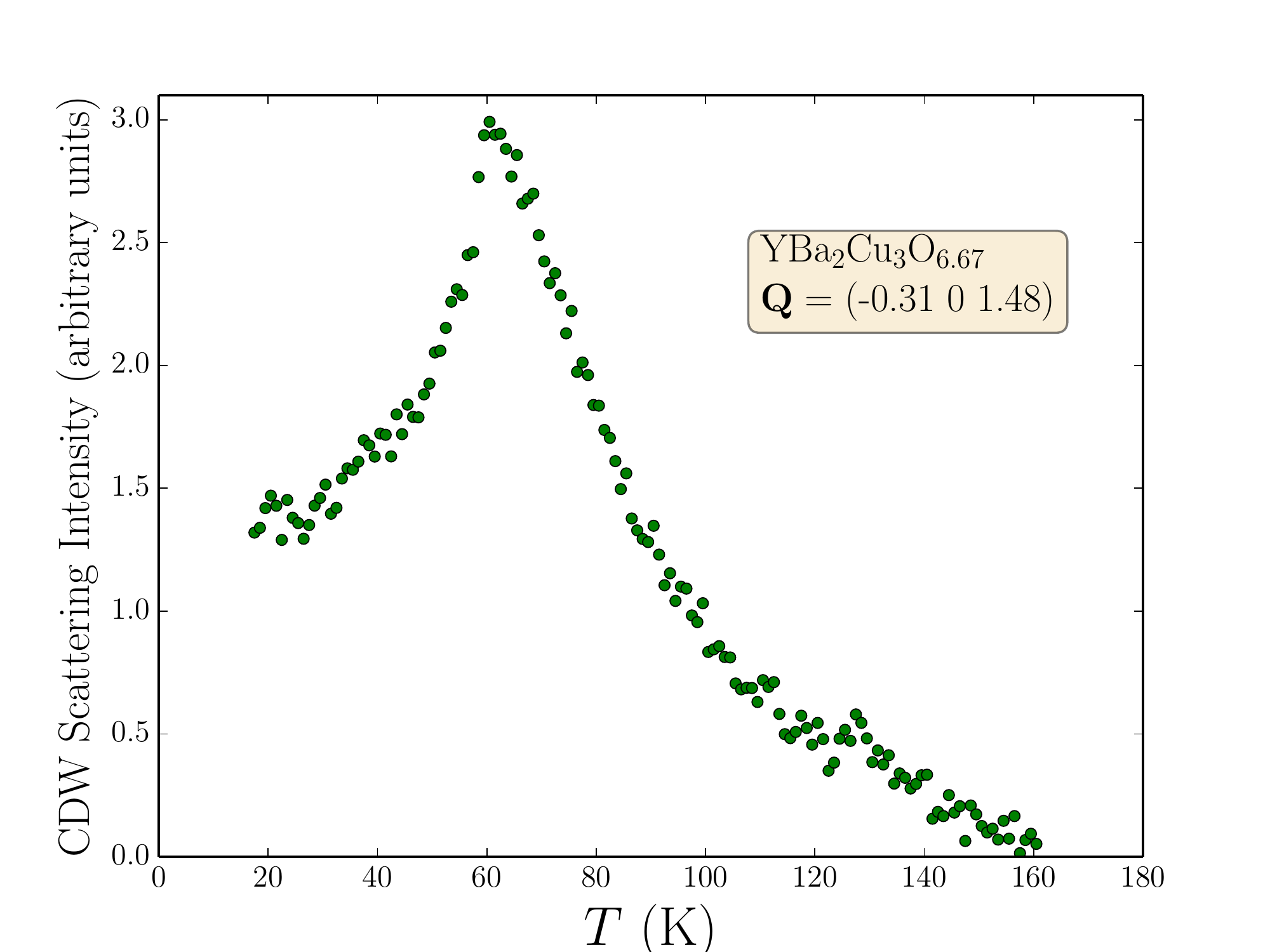}
\caption{The temperature dependence of the CDW scattering intensity at {\bf Q } = [-0.31 0 1.48] in YBa$_2$Cu$_3$O$_{6.67}$ measured by resonant x-ray scattering in Ref.~\cite{achkar2}. This sample has $T_c \approx 65.5$K. }
\label{fig:data}
\end{figure}

The X-ray scattering intensity \cite{achkar2} of YBa$_{2}$Cu$_3$O$_{6.67}$ 
at the incommensurate wavevectors $\bQ_x \approx (0.31,0)$ or $\bQ_y \approx (0,0.31)$, 
shown in Fig.~\ref{fig:data},
increases gradually below  $T \approx 200$K in a concave-upward shape until just above $T_c = 60$K.
One possibility is that this represents an order parameter of a broken symmetry, and the correlation length is arrested at a 
finite value by disorder; however, such order parameters invariably have a concave-downward shape.
The temperature range is also too wide to represent the precursor critical fluctuations of an ordering transition.
Indeed, there is no ordering transition below $T_c$, and, remarkably, the scattering intensity decreases below $T_c$ at a rate
similar to that of the rate of increase above $T_c$.

Instead, the increase in intensity between 200K and 60K is reminiscent of the 
classic measurement by Keimer {\em et al.} \cite{keimer2}, who observed a
gradual increase in the neutron scattering intensity
at the antiferromagnetic wavevector in the insulating antiferromagnet La$_2$CuO$_4$ between 550K and 350K \cite{neellro}. 
This increase was explained by the classical thermal, angular fluctuations
of the 3-component antiferromagnetic order parameter in $d=2$ spatial dimensions \cite{chn}.
Indeed, this is a special case of a general
result of Polyakov \cite{polyakov} who showed that order parameters with $N \geq 3$ components are dominated by angular fluctuations
in $d=2$; here, we will exploit the $N=6$ case to describe X-ray scattering in the pseudogap of YBa$_{2}$Cu$_3$O$_{6.67}$.

Previous work \cite{ssdemler} used a Landau theory framework \cite{zke} to describe
competition between superconductivity and charge density wave order \cite{concepts,rmp}. 
The Landau theory introduces a complex field $\Psi (\br)$ to represent the superconductivity, and two complex fields $\Phi_{x,y} (\br)$ to represent
the charge order. The latter can represent modulations at the incommensurate
wavevectors $\bQ_{x,y}$ in not only the site charge density, but also modulations in
bond variables associated with a pair of sites \cite{rmp,rolando}; nevertheless, we will refer to it simply as ``charge'' order. 
The free energy is restricted by 3 distinct U(1) symmetries:
charge conservation, translations in $x$, and translations in $y$, which rotate the phases of $\Psi$, $\Phi_x$, and $\Phi_y$ respectively.
There are also the discrete symmetries of time-reversal and the square lattice point group, and these lead to the following
form of the Landau free energy density (we ignore possible anisotropies in the spatial derivative terms):
\bea
F &=& |\nabla \Psi|^2 + s_1 |\Psi|^2 + u_1 |\Psi|^4 
 + |\nabla \Phi_x|^2 + |\nabla \Phi_y|^2 + s_2 
\left( | \Phi_x|^2 + | \Phi_y|^2 \right)\nonumber \\ &~&  + u_2 
\left( | \Phi_x|^2 + | \Phi_y|^2 \right)^2 + w \left( |\Phi_x|^4 + |\Phi_y|^4 \right) + v |\Psi|^2 \left(| \Phi_x|^2 + | \Phi_y|^2 \right)
\label{F}
\eea
The earlier analysis \cite{ssdemler} considered ``phase'' and ``vortex'' fluctuations of only the superconducting order, $\Psi$, and then
assumed that the charge order amplitude was proportional to $- v \left \langle |\Psi|^2 \right\rangle$, where $v>0$ is the competing order coupling: this analysis found a small decrease in charge order with decreasing $T$, but did not find a prominent peak near $T_c$. Here, we shall provide a theory which is non-perturbative
in $v$, and which includes the thermal fluctuations of both $\Psi$ and $\Phi_{x,y}$ self-consistently, and applies over a wide range of temperatures.

\begin{figure}
  \centering
  \includegraphics[width=3.1in]{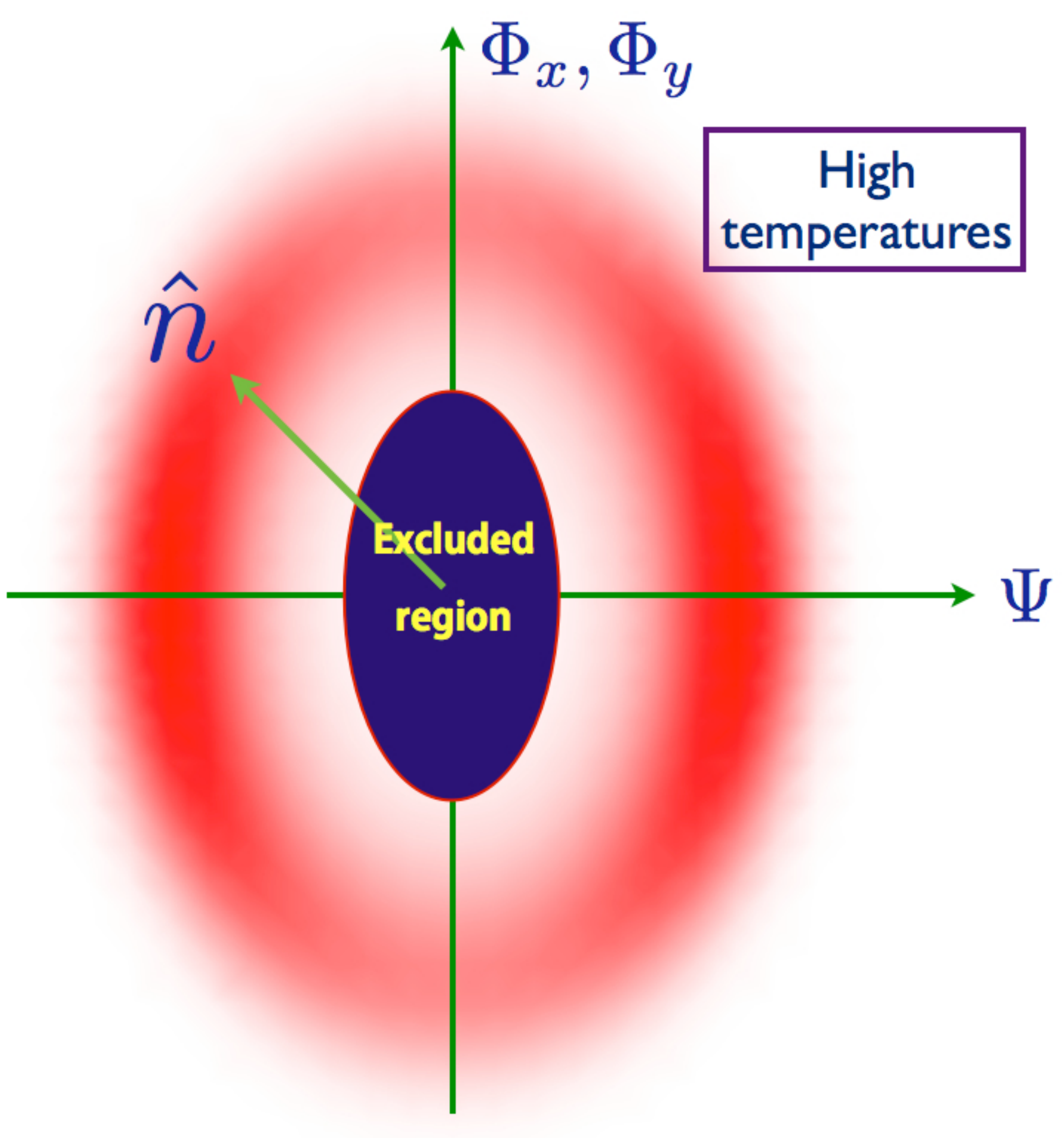}\\~\\~\\
  \includegraphics[width=3.1in]{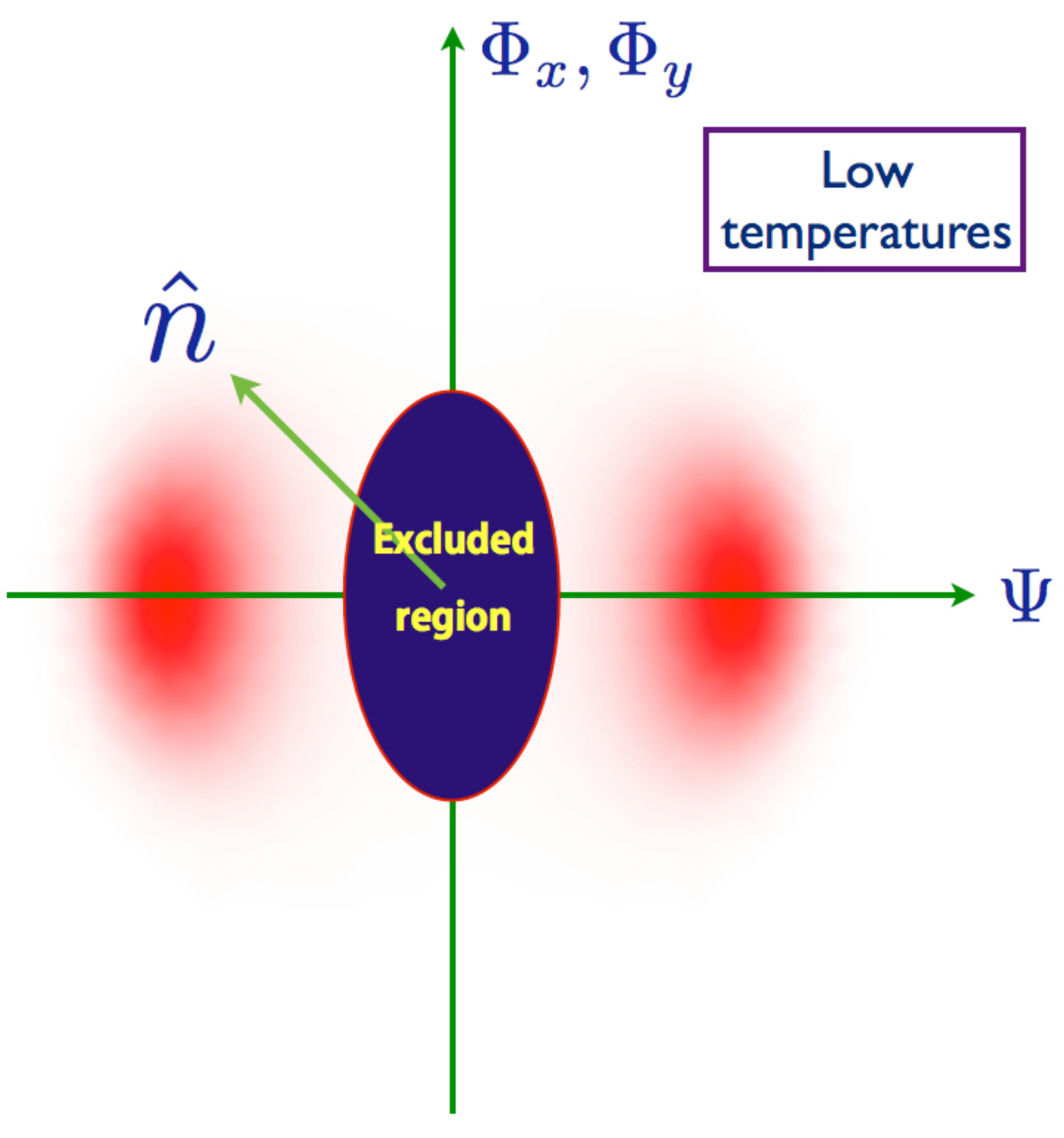}
  \caption{Schematic of the structure of fluctuations of $\mathcal{Z}$ in a 6-dimensional space representing the complex superconducting order, $\Psi$,
  and the complex charge orders $\Phi_{x,y}$. The red shading represents the probability that the values of $\Psi$, $\Phi_{x,y}$
  take particular values.
  At high $T$, all angles are explored, while at low $T$ below $T_c$, for $g>0$, the order lies mainly along
  the equator in the plane representing $\Psi$.
  }
  \label{fig:angular}
\end{figure}

Our starting assumption is that it is always preferable for the electronic Fermi surface to locally acquire some type of order, and so the 
origin of the 6-dimensional space defined by $(\Psi, \Phi_x, \Phi_y)$  should be excluded; see Fig.~\ref{fig:angular}.
For each radial direction in this 6-dimensional space, we can label the optimal state by a unit vector $n_{\alpha}$ ($\alpha = 1 \ldots 6$)
with $\Psi \propto n_1 + i n_2$, $\Phi_x \propto n_3 + i n_4$, and $\Phi_y \propto n_5 + i n_6$. 
We will neglect amplitude fluctuations along the radial direction and focus
solely on the angular fluctuations; no assumptions of an approximate O(6) symmetry are made a priori.
So we 
introduce a partition function for angular fluctuations of $n_\alpha$, with all terms allowed by the symmetries noted earlier:
\bea
\mathcal{Z} &=&  \int \mathcal{D} n_{\alpha} (\br) \, \delta \left(\sum_{\alpha=1}^6 n_{\alpha}^2 (\br) - 1 \right)  
\exp \Biggl( - \frac{\rho_s}{2 T}\int d^2 r \Biggl[  \sum_{\alpha=1}^2 (\nabla n_{\alpha})^2  
+ \lambda \sum_{\alpha=3}^6 (\nabla n_{ \alpha})^2 \nonumber \\
&~& \quad \quad \quad \quad \quad \quad + \, g \sum_{\alpha=3}^6 n_{\alpha}^2  +w 
\left[\left( n_{3}^2 + n_{4}^2 \right)^2 + \left( n_{5}^2 + n_{6}^2 \right)^2 \right] \Biggr]
\Biggr). \label{Z}
\eea
The couplings $\rho_s$ and $\rho_s \lambda$ are the helicity moduli for spatial variations of the superconducting and charge orders respectively.
The coupling $g$ measure the relative energetic cost of ordering between the superconducting and charge order directions; this is most relevant term which breaks the O(6) symmetry present for $\lambda=1$, $g=0$, $w=0$  to O(4)$\times$O(2) symmetry. Finally $w$ imposes the square lattice 
point group symmetry on the charge order: for $w<0$ the charge is uni-directional with only one of $\Phi_x$ or $\Phi_y$ non-zero, while for $w>0$ the
charge ordering is bi-directional. The final symmetry of $\mathcal{Z}$ is 
O(2)$\times$O(2)$\times$O(2)$\rtimes \mathbb{Z}_2$, where the 3 O(2)'s are enlarged by discrete symmetries from the 3 U(1)'s noted earlier, and the $\mathbb{Z}_2$ represents the 90$^\circ$ spatial rotation symmetry, whose 
spontaneous breaking is measured by the Ising-nematic order \cite{KFE98}
$m= |\Phi_x|^2 - |\Phi_y|^2$.

The enhanced symmetries of $\mathcal{Z}$ at $\lambda=1$, $g=0$, $w=0$ include 
two SO(4) rotation symmetries between $d$-wave superconductivity and incommensurate $d$-wave bond order 
 that emerge at low energies 
in the vicinity of a generic quantum critical point for the onset of antiferromagnetism in a metal \cite{metlitski} (but with charge order $\bQ$'s along the $(1,\pm 1)$ directions); a non-linear
sigma model of this theory was developed by Efetov {\em et al.} \cite{pepin} and applied to the phase diagram in a magnetic field \cite{pepin2}. 
It was also argued \cite{rolando} that these symmetries 
can be viewed as remnants of the
SU(2) pseudospin gauge invariances of Mott insulators \cite{affleck,dagotto,leewen}, 
when extended to metals with a strong local antiferromagnetic exchange coupling. And we also note the similarity to the
SO(5) non-linear sigma model of competing orders \cite{so5}, which has antiferromagnetism, rather than charge order, competing with superconductivity.

A crucial feature of our analysis of $\mathcal{Z}$ is that the couplings $\rho_s$, $g$, $\lambda$, and $w$ are assumed to be $T$-independent. 
The dependence on absolute temperature
arises only from the Boltzmann $1/T$ factor in $\mathcal{Z}$, and this strongly constrains our fits to the experimental data. This feature ensures our
restriction to angular and classical fluctuations in the order parameter space.

We computed the properties of $\mathcal{Z}$ using a classical Monte Carlo simulation. This was performed using the Wolff cluster
algorithm, after the continuum theory was discretized on a square lattice of spacing $a$. This lattice is not related to the underlying square lattice of Cu atoms in the cuprates; instead, it is just a convenient ultraviolet regularization of the continuum theory, and we don't expect 
our results to be sensitive to the particular regularization chosen. All length scales in our results will be proportional to
the value of $a$, and the value of $a$ has to be ultimately determined by matching one of them to experiments.
We performed simulations on lattice sizes up to 72$\times$72, and were able to control all finite size effects.

We also performed a $1/N$ expansion on a generalized model with $N$ components of $n_\alpha$, as described in the supplement.
It was found to be quite accurate for the charge order correlations, but does not properly describe the superconducting
correlations near $T_c$ and below.

\begin{figure}
\centering
\includegraphics[width=4.5in]{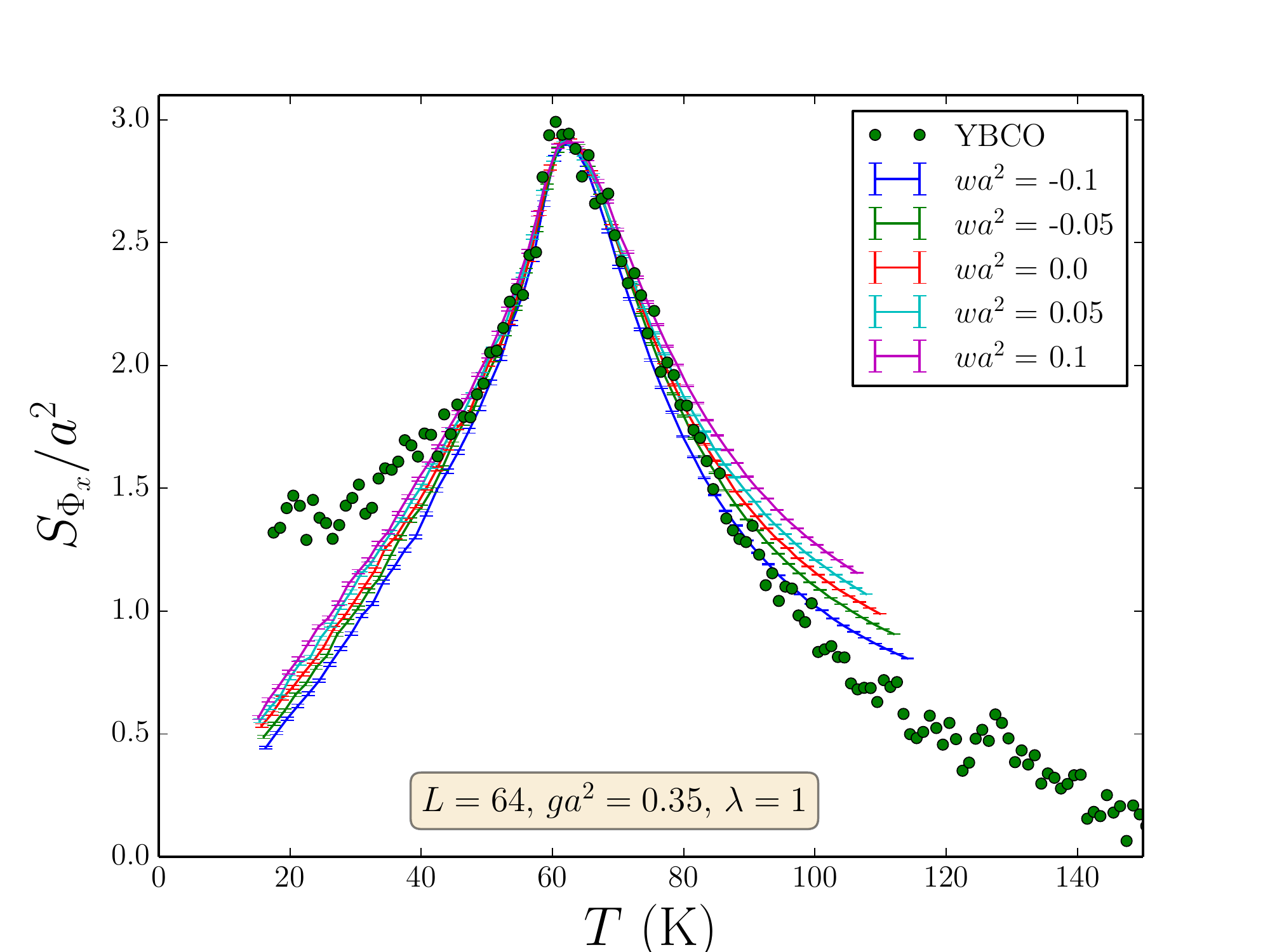}\\~\\~\\
\includegraphics[width=4.5in]{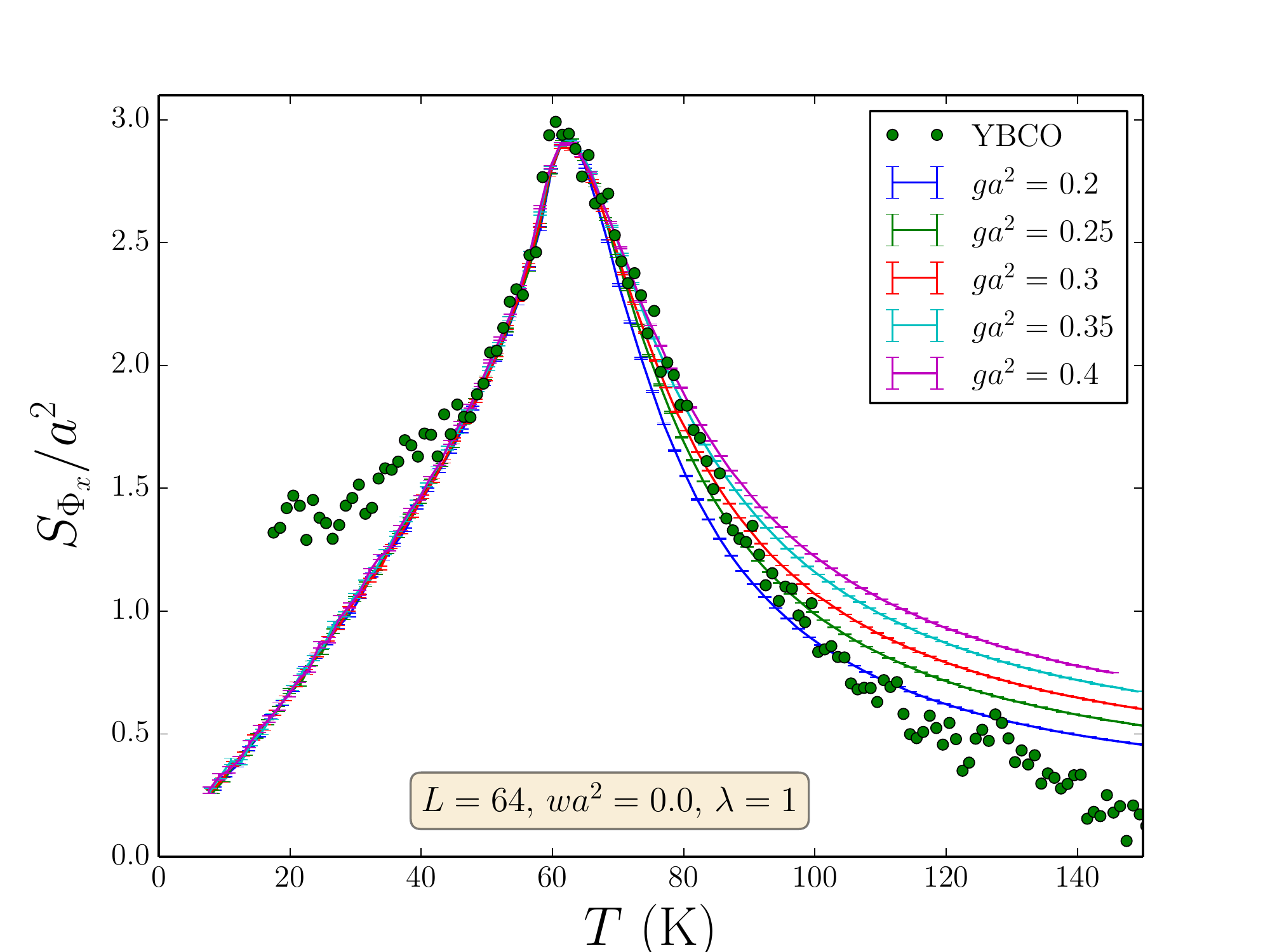}
\caption{Comparison of the X-ray data to Monte Carlo simulations of $\mathcal{Z}$.  Both axes measure dimensionless quantities.
For each set of values of $ga^2$, $wa^2$ and $\lambda$, there were 2 fitting parameters.
The value of $\rho_s$ was determined for each data set so that the peak
positions match: this is equivalent to a rescaling (but not shifting) of the $T$-axis, and does not determine the peak width or shape. For $ga^2=0.30$ and $wa^2 = 0.0$ we have $\rho_s = 160$K.
The height of the experimental data was rescaled to make the peak heights match.
}
\label{fig:sphi}
\end{figure}

\begin{figure}
\centering
\includegraphics[width=6in]{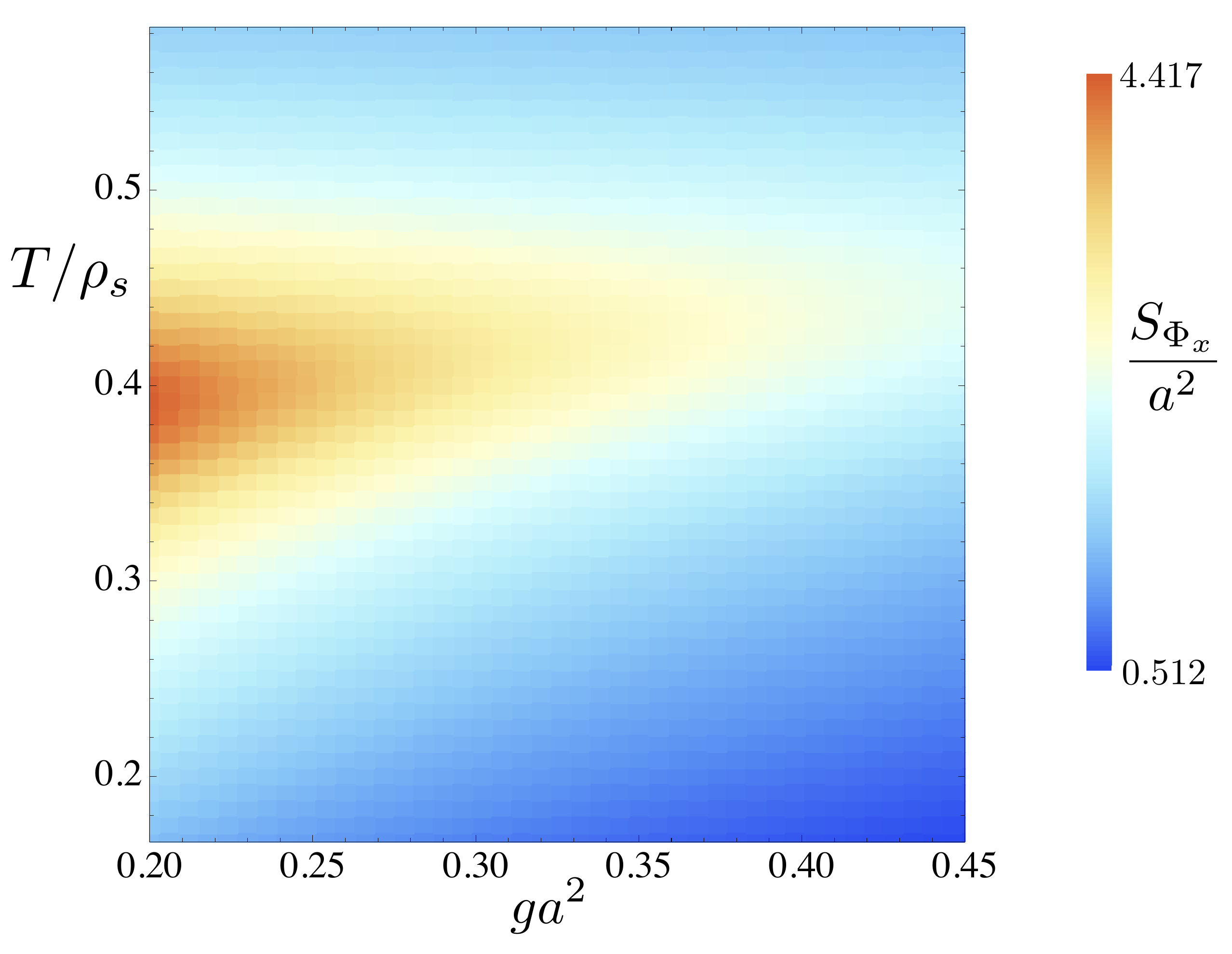}
\caption{Density plot of $S_{\Phi_x}$ as a function of $g a^2$ and $T/\rho_s$, for $\lambda = 1$
and $w a^2 = 0.1$ at order $1/N$ in the large $N$ expansion.}
\label{fig:sphi2}
\end{figure}

Our Monte Carlo results for the charge order correlations are shown in Fig.~\ref{fig:sphi}.
We computed the structure factor
\beq
S_{\Phi_x} (p) = \int d^2 r \sum_{\alpha=3}^4 \left\langle n_\alpha ( \br ) n_\alpha (0) \right\rangle e^{i \bp \cdot \br};
\eeq
in the X-ray experiments, $p$ represents deviations from the wavevectors $\pm \bQ_x$, $\pm \bQ_y$.
We show the values of $S_{\Phi_x} \equiv S_{\Phi_x} (p=0)$ for a variety of parameters. At high $T$, we have regime of increasing
$S_{\Phi_x}$ with decreasing $T$, as the correlation length of both the superconductivity and charge order increases, and the order parameter fluctuates
over all 6 directions (see Fig.~\ref{fig:angular}). At low $T$, there is onset of superconducting order, and $S_{\Phi_x}$ decreases
with decreasing $T$, as the order parameter becomes confined to the $\Psi$ plane. In Fig.~\ref{fig:sphi}, we fit the position of the peak
in $S_{\Phi_x}$ by choosing the value of $\rho_s$, and adjusted the vertical 
scale so that the peak height also coincides. Note that we are not allowed to shift the horizontal axis, as $T$ is predetermined. The peak width and shape
is {\em not\/} adjustable and is determined by the theory; so it can be used to fix the values of the dimensionless parameters $g a^2$, $w a^2$, and $\lambda$.
It is evident that the theory naturally reproduces the experimental curve, including the rate of decrease of charge order on both sides
of the peak, for a range of parameter values. Another view of $S_{\Phi_x}$ is in Fig.~\ref{fig:sphi2}, where we present results
of the $1/N$ expansion.

Note that there are differences between the experiment and theory in Fig.~\ref{fig:sphi}, 
both at very low and very high $T$. However, the deviations are 
in the expected directions. At low $T$, in the present classical theory without randomness, $S_{\Phi_x}$ vanishes as $T \rightarrow 0$; however, pinning of the charge order by impurities is likely responsible for the observed $S_{\Phi_x}$ by 
impurities \cite{tacon,laimei}.
At high $T$, our assumption of a $T$-independent bare $\rho_s$ starts failing at $T \approx 2 T_c$, and we expect a crossover
to a theory with significant amplitude fluctuations and smaller $S_{\Phi_x}$ with increasing $T$.

\begin{figure}
\centering
\includegraphics[width=4in]{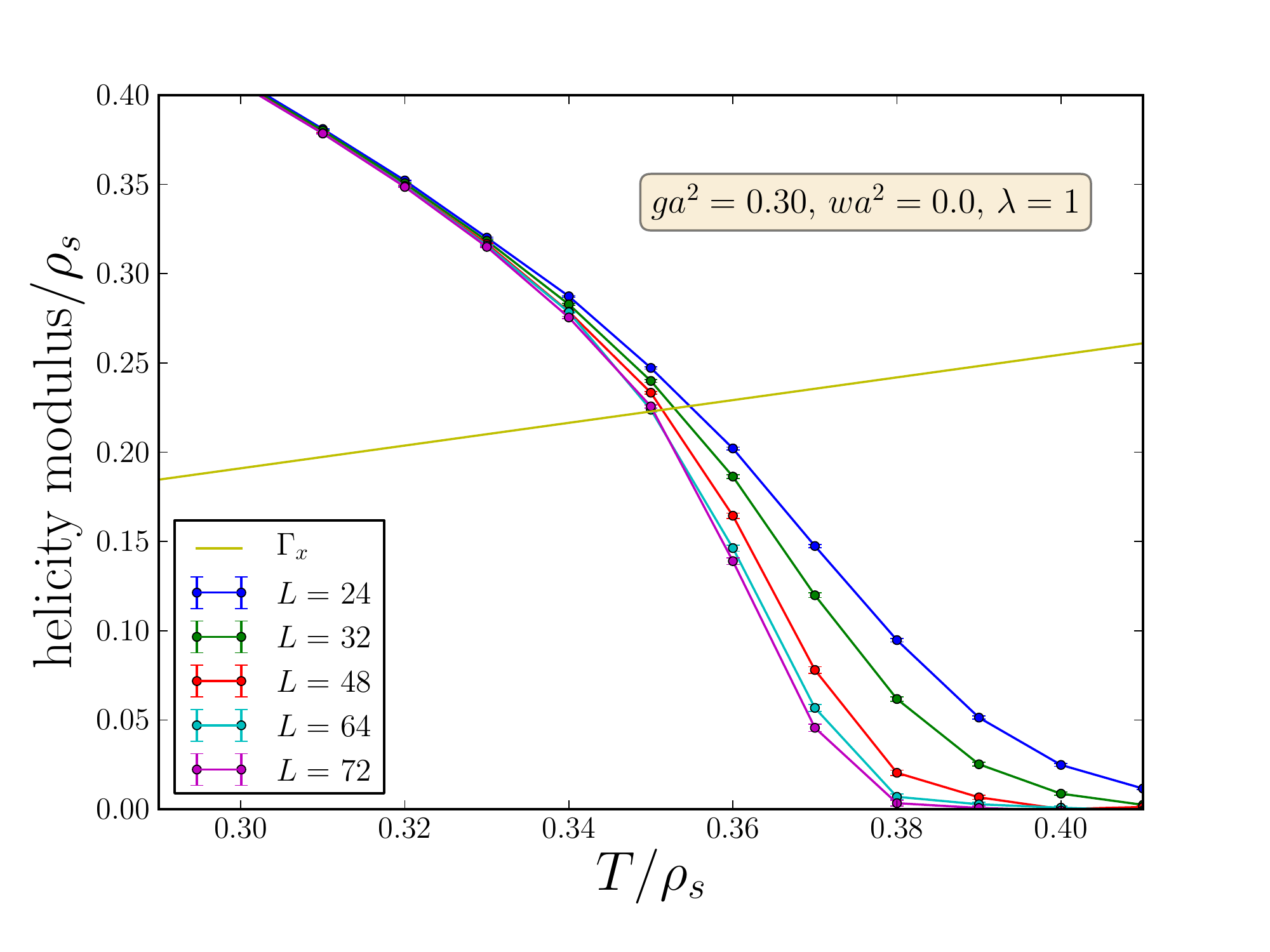}\\~\\~\\
\includegraphics[width=4in]{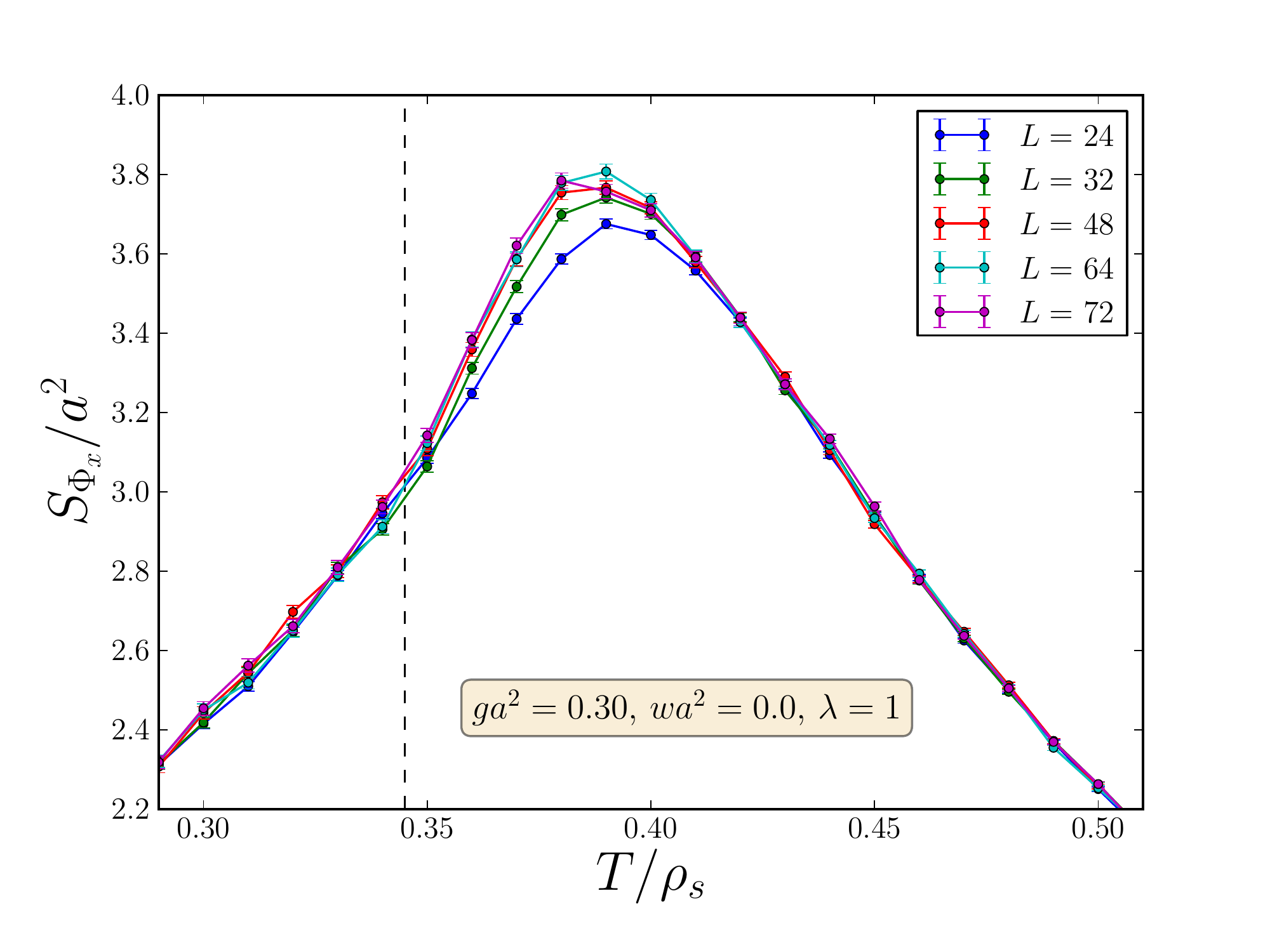}
\caption{Top: Monte Carlo results for the helicity modulus, measured in the $x$-direction. Note that $\rho_s$ is the helicity modulus
at $T=0$.
We also plot $\Gamma_x = (2/\pi) T/\rho_s$, and use the relation helicity modulus $=(2/\pi) T_c$ \cite{helicityjump}
to determine the Kosterlitz-Thouless temperature for each system size $L$.  A finite-size scaling analysis estimates $T_{c}/\rho_s \approx 0.345$ for these parameters.  Bottom: The structure factor, showing a peak at around $T/\rho_s=0.39$.  The 
Kosterlitz-Thouless temperature, $T_c$, is marked with a vertical dashed line. The prediction of Ref.~\cite{ssdemler} of increasing charge order with increasing temperature applies in the immediate vicinity of $T_c$, to the left of the peak.
}
\label{fig:helicity}
\end{figure}

Next, we examined the superconducting correlations by measuring the associated helicity modulus. As shown in Fig.~\ref{fig:helicity}, this allows
us to determine $T_c$ by comparing against the expected universal jump \cite{helicityjump}. 
We find a $T_c$ below the peak in $S_{\Phi_x}$.
This is consistent with the arguments in Ref.~\cite{ssdemler}, which predicted a monotonic decrease in charge order through $T_c$: evidently
their computations only apply in a narrow window about $T_c$. We note that we have
 not accounted for inter-layer couplings in our two-dimensional theory:
 these can raise $T_c$ and yield a cusp-like singularity in the charge order at $T_c$.

One of the fundamental aspects of our theory is that the {\em same\/} set of parameters used above to describe X-ray scattering experiments,
also predict the strength of superconducting fluctuations above $T_c$. The latter are detectable in diamagnetism measurements,
and indeed YBa$_{2}$Cu$_3$O$_{6+x}$ shows significant fluctuation diamagnetism \cite{ong,ybcomag} 
over the range of temperatures that X-ray experiments
measure charge order fluctuations. We compute the diamagnetic susceptibility in the $N=\infty$ theory in the supplement.
Such a theory has effectively Gaussian superconducting fluctuations, albeit with a $T$ dependence of the superconducting
coherence length which is different from the standard Landau-Ginzburg form \cite{larkin}.
An absolute comparison of this theory with the observations \cite{ybcomag} yields the value of $a$, 
which is found differ by about 33\% from the value 
obtained from the charge order correlation length. Considering the simplicity of the $N=\infty$ theory, 
the possible differences in the X-ray and diamagnetism samples, and the absence of fitting to determine
$\lambda$ and $w$, this result is encouraging. 
For a sharper comparison, we need to study the crossover into a vortex dominated
regime \cite{vortex1,vortex2,vortex3}:
its description requires Monte Carlo study in our theory, which is in progress.
Eventually, with a complete study which also includes the effects of disorder, and more detailed
measurements of charge order and superconducting correlations on the same sample, we expect to be able to more tightly constrain the values
of $ga^2$, $wa^2$, $\lambda$, and $a$.

Placing this work in a broader context, although we have only applied the theory to a doping where charge order is 
most pronounced, we argue that it is characteristic of the entire pseudogap phase.  The dominant paradigms for the pseudogap 
have been phase fluctuating superconductivity \cite{emery} and competing order \cite{concepts,rmp}, with experiments providing 
merit to both descriptions \cite{Corson,Xu,keimer,chang,hawthorn,julienvortex}.  This work unifies these paradigms in a 
single multi-component order parameter which provides a natural description of the X-ray and diamagnetism data. 
Computations of the influence of fluctuating superconductivity on photoemission 
spectra \cite{mohit} should now be extended to include all components of our order parameter.
Our model is also linked to theories of metals with antiferromagnetic spin 
fluctuations \cite{metlitski,pepin}:
with decreasing doping, there is a zero field quantum critical point to the onset of antiferromagnetic order \cite{julien2}, and this indicates that our present model will have to be extended to explicitly include spin fluctuations \cite{keimer3} to apply at such densities. 

{\textit{Acknowledgments}}. 
 We thank A.~Chubukov, A.~Georges, M.-H.~Julien, B.~Keimer, 
S.~Kivelson, J.~Orenstein, and A.~Yacoby for useful discussions.
This research
was supported by the NSF under Grant DMR-1103860 and the Natural Sciences and Engineering Research Council of Canada.
This research was also supported in part by Perimeter Institute for
Theoretical Physics; research at Perimeter Institute is supported by the
Government of Canada through Industry Canada and by the Province of
Ontario through the Ministry of Research and Innovation.

\appendix

\section{Large $N$ expansion}

We carried out the large $N$ expansion of the partition function $\mathcal{Z}$ by generalizing it to a model with
a $N$-component unit vector $n_{\alpha}$, in which the O($N$) symmetry breaks down to\\ 
O($N/3$)$\times$O($N/3$)$\times$O($N/3$)$\rtimes \mathbb{Z}_2$. The action for such a model is 
\bea
\mathcal{S} &=& \frac{\rho_s}{2 T}\int d^2 r \left\{  \sum_{\alpha=1}^{N/3} (\nabla n_{\alpha})^2  
+ \lambda \sum_{\alpha=N/3+1}^N (\nabla n_{ \alpha})^2  + \, g \sum_{\alpha=N/3+1}^N n_{\alpha}^2 \right. \nn
&~&~~~~~~\quad \quad \quad  \left.+w 
\left[\left( \sum_{\alpha = N/3+1}^{2N/3} n_{\alpha}^2 \right)^2 + \left( \sum_{\alpha=2N/3+1}^N n_{\alpha}^2  \right)^2 \right] \right\}.
\eea
The large $N$ expansion proceeds by a standard method \cite{brezin}, and requires that 
\beq
T =t/N, 
\eeq
with $t$ of order unity. 
We introduce an auxilliary field $\sigma$ to impose the unit length constraint, and two fields $\phi_{x,y}$ 
which decouple the quartic terms. In this manner we obtain
\bea
\mathcal{S} &=& \frac{N\rho_s}{2 t}\int d^2 r \left\{  \sum_{\alpha=1}^{N/3} (\nabla n_{\alpha})^2  
+ \lambda \sum_{\alpha=N/3+1}^N (\nabla n_{ \alpha})^2  + \, g \sum_{\alpha=N/3+1}^N n_{\alpha}^2 \right. \nn
&~& \left. + i \sigma \left( \sum_{\alpha=1}^N n_\alpha^2 - 1 \right) + \frac{\phi_x^2 + \phi_y^2}{4w} + i \phi_x \sum_{\alpha=N/3+1}^{2N/3} n_\alpha^2 +  i \phi_y \sum_{\alpha=2N/3+1}^{N} n_\alpha^2
\right\}.
\eea

In the $N=\infty$ limit, we can integrate out the $n_\alpha$, and the auxilliary fields are all fixed at their saddle-point values $i \sigma = \overline{\sigma}$, $i \phi_{x,y} = \overline{\phi}_{x,y}$
which are determined by the saddle point equations
\bea
\frac{\rho_s}{t} &=& \frac{1}{3} \int_{\bp} \left[ \frac{1}{p^2 + \overline{\sigma}} + \frac{1}{\lambda p^2 + \overline{\sigma} + g + \overline{\phi}_x} + \frac{1}{\lambda p^2 + \overline{\sigma} + g + \overline{\phi}_y} \right] \nonumber \\
 \overline{\phi}_x &=& \frac{2 w t}{3 \rho_s} \int_{\bp} \frac{1}{\lambda p^2 + \overline{\sigma} + g + \overline{\phi}_x} \nonumber \\
 \overline{\phi}_y &=& \frac{2 w t}{3 \rho_s} \int_{\bp} \frac{1}{\lambda p^2 + \overline{\sigma} + g + \overline{\phi}_y} 
\eea
where $\int_{\bp} \equiv \int d^2 p/(4 \pi^2)$.
The optimum solution minimizes the free energy density, which is given by
\beq
F = \frac{t}{6} \int_{\bp}\ln  \left[(p^2 + \overline{\sigma})  ( \lambda p^2 + \overline{\sigma} + g + \overline{\phi}_x)  ( \lambda p^2 + \overline{\sigma} + g + \overline{\phi}_y) \right] - \frac{\rho_s \overline{\sigma}}{2} - \frac{\rho_s (\overline{\phi}_x^2 + \overline{\phi}_y^2)}{8w}
\eeq
A solution with $\overline{\phi}_x \neq \overline{\phi}_y$ breaks Ising-nematic symmetry, and this happens at sufficiently low temperatures for $w<-g$ 
and $g>0$, or for $w<0$ and $g<0$.
The momentum-dependent structure factors of the $\Psi = (n_1, \ldots, n_{N/3})$, 
$\Phi_x= (n_{N/3+1}, \ldots, n_{2N/3})$, and $\Phi_y = (n_{2N/3+1}, \ldots , n_N)$ correlators are
\bea
S_\Psi (p) &=& \frac{t/(3\rho_s)}{p^2 + \overline{\sigma}} \nn 
S_{\Phi_x} (p) &=& \frac{t/(3\rho_s)}{\lambda p^2 + \overline{\sigma} + g + \overline{\phi}_x} \nn
S_{\Phi_y} (p) &=& \frac{t/(3\rho_s)}{\lambda p^2 + \overline{\sigma} + g + \overline{\phi}_y} . \label{sninf}
\eea

\begin{figure}
  \centering
  \includegraphics[width=6in]{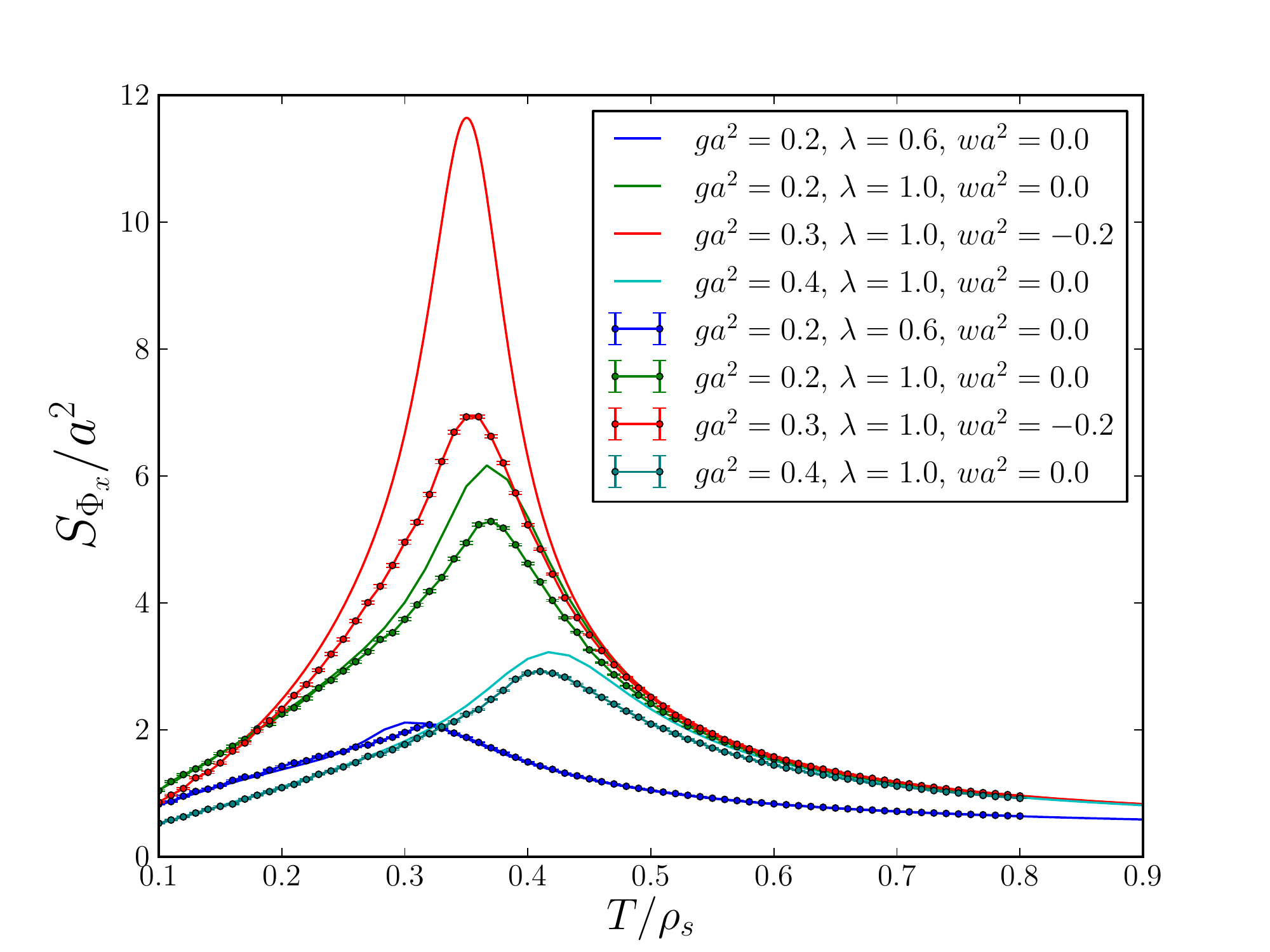}
  \caption{Comparison of the charge order structure factor as obtained from the large $N$ expansion at order $1/N$, with the computations
  of the Monte Carlo for the same parameters, and size $L=32$.  Large $N$ calculations are solid lines, and Monte Carlo data is plotted as circles with statistical error bars.
  }
  \label{fig:comparison}
\end{figure}

For the $1/N$ corrections, we need to include fluctuations of $\sigma, \phi_{x,y}$ about their saddle point values. 
See Ref.~\cite{podolsky} for details on a similar computation in a different context.
The propagators of these fields
are expressed in terms of `polarization functions' which are given by
\bea
\Pi (p,\overline{\sigma}) &=&  \frac{1}{3} \int_\bq  \biggl[ \frac{1}{(q^2+\overline{\sigma})((\bp + \bq)^2 + \overline{\sigma})} + \frac{1}{(\lambda q^2+\overline{\sigma}+g + \overline{\phi}_x)(\lambda (\bp + \bq)^2 + \overline{\sigma}+g + \overline{\phi}_x)} \nn
&~& \quad\quad\quad\quad +\frac{1}{(\lambda q^2+\overline{\sigma}+g + \overline{\phi}_y)(\lambda (\bp + \bq)^2 + \overline{\sigma}+g + \overline{\phi}_y)}  \biggr] \nn
\Pi_x (p,\overline{\sigma}) &=&  \frac{\rho_s}{2wt} + \frac{1}{3} \int_\bq  \frac{1}{(\lambda q^2+\overline{\sigma}+g + \overline{\phi}_x)(\lambda (\bp + \bq)^2 + \overline{\sigma}+g + \overline{\phi}_x)} \nn
\Pi_y (p,\overline{\sigma}) &=&  \frac{\rho_s}{2wt} + \frac{1}{3} \int_\bq  \frac{1}{(\lambda q^2+\overline{\sigma}+g + \overline{\phi}_y)(\lambda (\bp + \bq)^2 + \overline{\sigma}+g + \overline{\phi}_y)}  \label{eq:pol}
\eea
Then after including self-energy corrections in the $n_\alpha$ propagators, we obtain the $1/N$ corrections to Eq.~(\ref{sninf}):
\bea
 \frac{t}{3 \rho_s} S^{-1}_\Psi (p) &=& p^2 + \overline{\sigma}  + \frac{1}{N} \frac{1}{\Pi (0,\overline{\sigma})} \int_{\bq} \frac{1}{\Pi (q,\overline{\sigma})}  
\left[ \frac{d \Pi (q,\overline{\sigma})}{d\overline{\sigma}} + \frac{2 \Pi (0,\overline{\sigma})}{((\bp + \bq)^2 + \overline{\sigma})} \right]  \nn
&~&~+ \frac{1}{N} \frac{1}{\Pi (0,\overline{\sigma})} \int_{\bq} \frac{1}{\Pi_x (q,\overline{\sigma})}  
 \frac{d \Pi_x (q,\overline{\sigma})}{d\overline{\sigma}}  + \frac{1}{N} \frac{1}{\Pi (0,\overline{\sigma})} \int_{\bq} \frac{1}{\Pi_y (q,\overline{\sigma})}  
 \frac{d \Pi_y (q,\overline{\sigma})}{d\overline{\sigma}} \nn
 \frac{t}{3 \rho_s} S^{-1}_{\Phi_x} (p) &=& \lambda p^2 + \overline{\sigma}  +g+ \overline{\phi}_x + \frac{1}{N} \frac{1}{\Pi (0,\overline{\sigma})} \int_{\bq} \frac{1}{\Pi (q,\overline{\sigma})}  
\left[ \frac{d \Pi (q,\overline{\sigma})}{d\overline{\sigma}} + \frac{2 \Pi (0,\overline{\sigma})}{(\lambda (\bp + \bq)^2 + \overline{\sigma}+g + \overline{\phi}_x)} \right] \nn
&~&~+ \frac{1}{N} \frac{1}{\Pi (0,\overline{\sigma})} \int_{\bq} \frac{1}{\Pi_x (q,\overline{\sigma})}  
 \frac{d \Pi_x (q,\overline{\sigma})}{d\overline{\sigma}}  + \frac{1}{N} \frac{1}{\Pi (0,\overline{\sigma})} \int_{\bq} \frac{1}{\Pi_y (q,\overline{\sigma})}  
 \frac{d \Pi_y (q,\overline{\sigma})}{d\overline{\sigma}} \nn
 &~&~+ \frac{1}{N} \frac{1}{\Pi_x (0,\overline{\sigma})} \int_{\bq} \frac{1}{\Pi_x (q,\overline{\sigma})}  
\left[ \frac{d \Pi_x (q,\overline{\sigma})}{d\overline{\sigma}} + \frac{2 \Pi_x (0,\overline{\sigma})}{(\lambda (\bp + \bq)^2 + \overline{\sigma}+g + \overline{\phi}_x)} \right] 
\nn
&~&~+ \frac{1}{N} \frac{1}{\Pi_x (0,\overline{\sigma})} \int_{\bq} \frac{1}{\Pi (q,\overline{\sigma})}  
 \frac{d \Pi_x (q,\overline{\sigma})}{d\overline{\sigma}}
\nn
 \frac{t}{3 \rho_s} S^{-1}_{\Phi_y} (p) &=& \lambda p^2 + \overline{\sigma}  +g+ \overline{\phi}_y + \frac{1}{N} \frac{1}{\Pi (0,\overline{\sigma})} \int_{\bq} \frac{1}{\Pi (q,\overline{\sigma})}  
\left[ \frac{d \Pi (q,\overline{\sigma})}{d\overline{\sigma}} + \frac{2 \Pi (0,\overline{\sigma})}{(\lambda (\bp + \bq)^2 + \overline{\sigma}+g + \overline{\phi}_y)} \right] \nn
&~&~+ \frac{1}{N} \frac{1}{\Pi (0,\overline{\sigma})} \int_{\bq} \frac{1}{\Pi_x (q,\overline{\sigma})}  
 \frac{d \Pi_x (q,\overline{\sigma})}{d\overline{\sigma}}  + \frac{1}{N} \frac{1}{\Pi (0,\overline{\sigma})} \int_{\bq} \frac{1}{\Pi_y (q,\overline{\sigma})}  
 \frac{d \Pi_y (q,\overline{\sigma})}{d\overline{\sigma}} \nn
 &~&~+ \frac{1}{N} \frac{1}{\Pi_y (0,\overline{\sigma})} \int_{\bq} \frac{1}{\Pi_y (q,\overline{\sigma})}  
\left[ \frac{d \Pi_y (q,\overline{\sigma})}{d\overline{\sigma}} + \frac{2 \Pi_y (0,\overline{\sigma})}{(\lambda (\bp + \bq)^2 + \overline{\sigma}+g + \overline{\phi}_y)} \right] 
\nn
&~&~+ \frac{1}{N} \frac{1}{\Pi_y (0,\overline{\sigma})} \int_{\bq} \frac{1}{\Pi (q,\overline{\sigma})}  
 \frac{d \Pi_y (q,\overline{\sigma})}{d\overline{\sigma}}
 \label{res}
\eea
We evaluated these expressions numerically after regulating the theory on a square lattice with 
lattice spacing $a$. Operationally, this means
that we perform the replacement $p^2 \rightarrow (4 - 2 \cos (p_x a) - 2 \cos (p_y a) )/a^2$ in all propagators, and the $p_{x,y}$ integrals
extend from $-\pi/a$ to $\pi/a$. We show our results for the equal-time structure factor of the charge order correlations
$S_{\Phi_x} \equiv S_{\Phi_x} (p=0)$ in Fig.~\ref{fig:comparison}.
For the parameters for which results are shown, we found good convergence upon replacing each
integral by a discrete sum over 
200 points. It is evident that the $1/N$ expansion is quite accurate, except near the peaks.

\subsection{Ising-nematic correlations}

\begin{figure}
  \centering
  \includegraphics[width=6in]{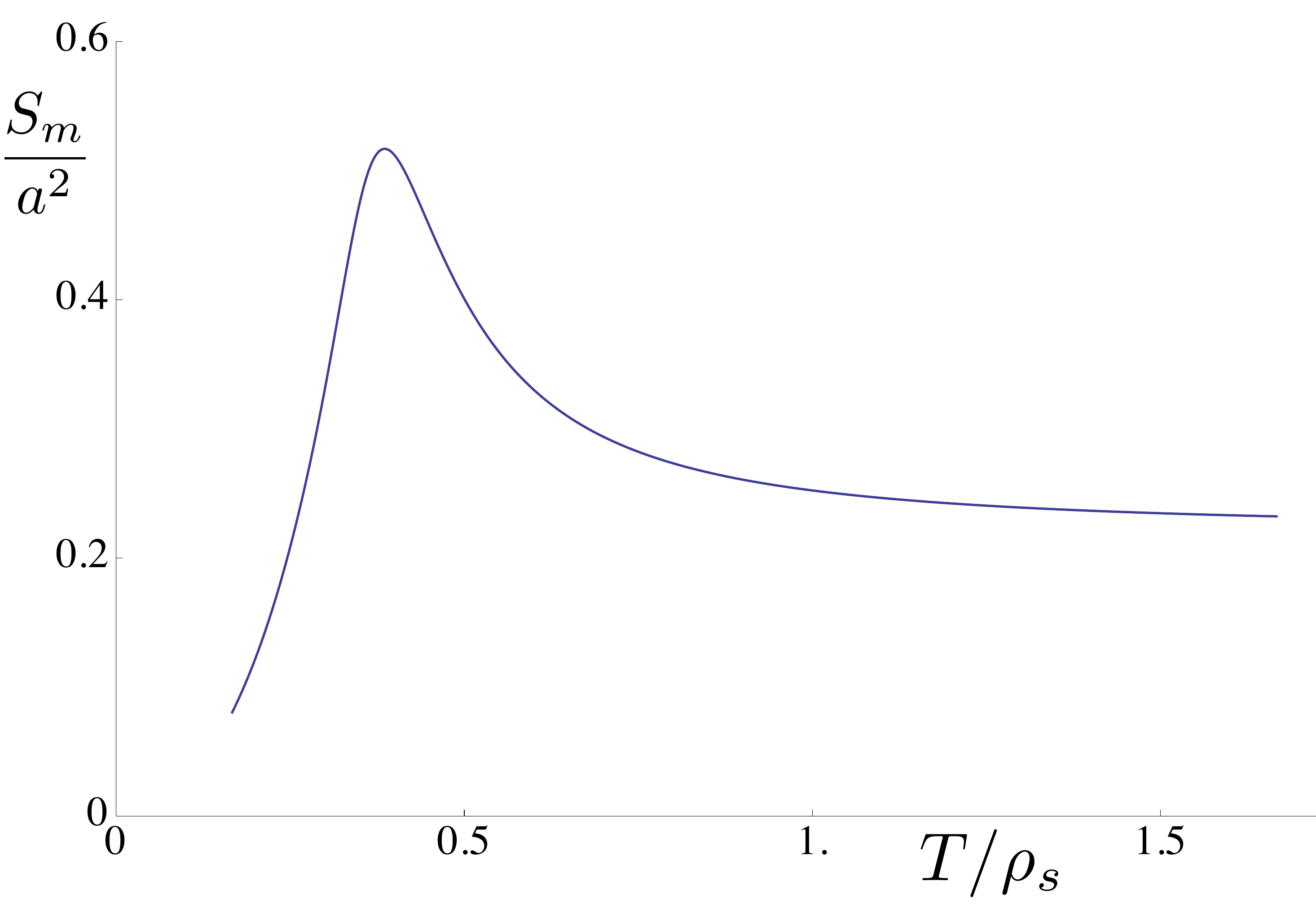}
  \caption{Ising-nematic structure factor,  as computed in the $N=\infty$ theory for $ga^2 = 0.3$, $\lambda=1$ and 
  $wa^2 = -0.2$. The corresponding charge ordering structure factor for these parameters is shown in Fig.~\ref{fig:comparison}.
  }
  \label{fig:nematic}
\end{figure}

We also computed the structure factor of the Ising-nematic order in the phase where Ising-nematic order is preserved. The Ising-nematic order
is $m = \sum_{\alpha=N/3+1}^{2N/3} n_\alpha^2 - \sum_{\alpha=2N/3 +1}^N n_\alpha^2$ and $S_m$ is its two-point correlator. We compute this
by including a source $J$ in the action $\mathcal{S} \rightarrow \mathcal{S} + \int d^2 r \, J \, m$. Then, after shifting the auxiliary fields
and integrating out the $n_\alpha$, we find that the effective action for the auxiliary fields maps via
\beq
\mathcal{S}[\sigma, \phi_x, \phi_y] \rightarrow \mathcal{S}[\sigma, \phi_x, \phi_y]  + \frac{i}{2 w} J \left( \phi_x - \phi_y \right) - \frac{t}{N \rho_s w} J^2
\eeq
By taking functional derivatives with respect to $J$, and then setting $J=0$, we can now relate the Ising structure factor to the 2-point
correlation of the auxiliary fields:
\beq
S_m (p) = \frac{2t}{N \rho_s w} - \frac{1}{4w^2} \int d^2 r\, e^{i \bp \cdot \br} \left\langle \left(\phi_x (\br) - \phi_y (\br) \right)
\left(\phi_x (0) - \phi_y (0) \right) \right\rangle
\eeq
At leading order in the $1/N$ expansion we can evaluate the correlator using the polarization functions in Eq.~(\ref{eq:pol}); because we are in the Ising-symmetric phase, $\Pi_x = \Pi_y$, and
\beq
N S_m (p) = \frac{2t}{w \rho_s} - \frac{1}{w^2 \Pi_x (p, \overline{\sigma})} = \frac{4 (t/\rho_s)^2 P(p)}{3 + 2 w (t/\rho_s) P(p)}
\eeq
where
\beq
P(p) = \int_\bq  \frac{1}{(\lambda q^2+\overline{\sigma}+g + \overline{\phi}_x)(\lambda (\bp + \bq)^2 + \overline{\sigma}+g + \overline{\phi}_x)}
\eeq
We show the $T$ dependence of $S_m \equiv S_m (p=0)$ in Fig.~\ref{fig:nematic} for a particular set of couplings.
\newpage

\subsection{Diamagnetic susceptibility}

We now compute the linear response to a magnetic field applied perpendicular to the layer in the $N=\infty$ theory.
We assume that the field only has an orbital coupling to the superconducting order. 
Here, we will carry out the computation explicitly with lattice regularization, on a square lattice of spacing $a$,
because we want to keep all expressions properly gauge-invariant.

At $N=\infty$ we can set $i \sigma = \overline{\sigma}$, and just treat the $\alpha = 1,2$ components of $n_\alpha$ as
Gaussian fields. Here, we normalize the complex superconducting order as $\widetilde{\Psi} = (n_1 + i n_{2})/\sqrt{2t/(N\rho_s)}$.
Then the part of the action that detects the presence of the magnetic field is 
\beq
\mathcal{S}_\Psi = - 
\sum_{\langle ij \rangle} \left( \widetilde{\Psi}_{i}^\ast \widetilde{\Psi}_{j} e^{i A_{ij}} + \mbox{c.c} \right) + \sum_i (4+ \overline{\sigma} a^2) |\widetilde{\Psi}_{i}|^2 \label{Spsi}
\eeq
where $A_{ij}$ is the Peierls phase from the applied field. 
The paramagnetic current is
\beq
J_i (\bq) = \frac{2}{a} \int \frac{d^2 k}{4 \pi^2} \widetilde{\Psi}^\ast (\bk + \bq/2) \widetilde{\Psi} (\bk - \bq/2) \sin (k_i a)
\eeq
So the 2-point current correlator, including the diamagnetic contribution, is
\bea
&& K_{ij} (\bq) = \left\langle J_i (\bq) J_j (-\bq) \right\rangle \nn
&&=  \frac{1}{a^2} \int \frac{d^2 k}{4 \pi^2} \frac{4\sin (k_i a) \sin (k_j a) }{
((4 - 2 \cos ((k_x + q_x/2)a) - 2 \cos ((k_y + q_y/2)a))/a^2 + \overline{\sigma})} \nn
&& \quad \quad \quad \quad \quad \quad\times \frac{1}{((4 - 2 \cos ((k_x - q_x/2)a) - 2 \cos ((k_y - q_y/2)a))/a^2 + \overline{\sigma})} \nn
&&~\quad \quad \quad ~ - \delta_{ij} \int \frac{d^2 k}{4 \pi^2} \frac{2 \cos (k_x a) }{((4 - 2 \cos (k_x a) - 2 \cos (k_y a))/a^2 + \overline{\sigma})}
\eea
This vanishes at $\bq = 0$ as expected by gauge invariance. For small $\bq$ we obtain
\beq
K_{ij} (\bq) = - (q^2 \delta_{ij} - q_i q_j) \frac{1}{a^4} \int \frac{d^2 k}{4 \pi^2} \frac{8 \sin^2 (k_x a) \sin^2 (k_y a) }{((4 - 2 \cos (k_x a) - 2 \cos (k_y a))/a^2 + \overline{\sigma})^4}
\eeq
For small $\overline{\sigma}$, the integral can be evaluated near $\bk=0$, and we obtain
\beq
K_{ij} (\bq) = - \frac{(q^2 \delta_{ij} - q_i q_j) }{12 \pi \overline{\sigma}}
\eeq
Restoring physical units, this implies that the magnetic susceptibility is 
\beq
\chi = - \frac{1}{s}\left( \frac{2e}{\hbar} \right)^2 \frac{k_B T}{12 \pi \overline{\sigma}} \label{chid}
\eeq
where $s$ is the interlayer spacing. This agrees precisely with the standard result \cite{larkin} in Eq.~(1) of Ref.~\cite{ybcomag},
after we observe from Eq.~(\ref{Spsi}) that $\overline{\sigma}$ is equal to 
$\xi_{ab}^{-2} (T)$, where $\xi_{ab} (T)$ is the superconducting coherence length.

\begin{figure}
  \centering
  \includegraphics[width=6in]{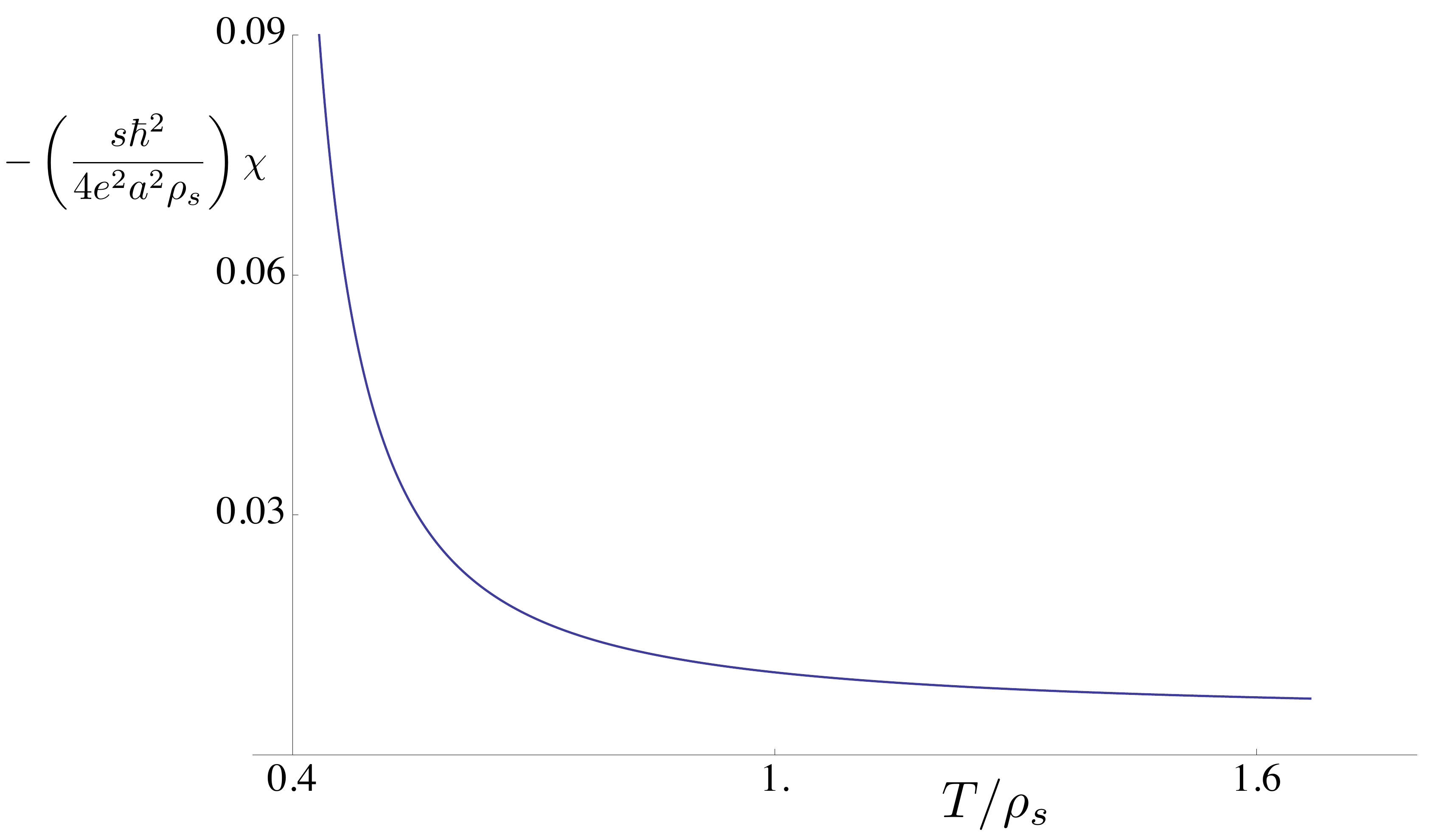}
  \caption{Diamagnetic susceptibility for the same set of parameters as in Fig.~\ref{fig:nematic}. The vertical axis is dimensionless,
  and its value in the $N=\infty$ theory is $(k_B T/\rho_s)/(12 \pi \overline{\sigma} a^2)$.
  }
  \label{fig:dia}
\end{figure}

We plot the $T$ dependence of $\chi$ in Fig.~\ref{fig:dia} for the same set of parameters used in Fig.~\ref{fig:nematic}.
We have only shown higher $T$ values because the large $N$ theory, which is effectively a Gaussian theory, is not reliable
close to the superconducting $T_c$. Note that the $T$ dependence of $\chi$ is similar to that in the observations \cite{ybcomag}.

For an absolute comparison, we note 
that for the diamagnetism 
data of Ref.~\cite{ybcomag}, a fit to the form in Eq.~(\ref{chid}) at 70 K for their UD57 sample yields \cite{cooper} a value
of $\overline{\sigma} \approx (39\, \mbox{\AA})^{-2}$.  For the $N=\infty$ theory, we use the results at the $T$ which has
the same ratio with position in the peak of the charge order, which is $k_B T/\rho_s = 0.44$; at this $T$, the $N=\infty$ theory results in
Fig.~\ref{fig:dia} yield $\overline{\sigma} a^2 = 0.14$.
Fitting this to the diamagnetic observations we obtain $a \approx 15 \, \mbox{\AA}$.

An independent estimate of $a$ can be obtained from the X-ray observations of Ref.~\cite{achkar2}, but it must be noted that these
are for a different sample. Their 
o-VIII sample has $T_c = 65.5$ K, and we use the charge order correlation length at the $T$ with the same ratio to $T_c$ as in the
diamagnetic data: this has $T=80$ K where $\xi_{\rm cdw} \approx  40\, \mbox{\AA}$. In our $N=\infty$ theory, for the same parameters as in Figs.~\ref{fig:nematic} and~\ref{fig:dia}, 
the charge order structure factor in Eq.~(\ref{sninf}) yields $\xi_{\rm cdw}^{-2} a^2 = (\overline{\sigma} + g + \overline{\phi}_x) a^2 = (\overline{\sigma} + g + \overline{\phi}_y) a^2 = 0.32$
at $k_B T/\rho_s = 0.44$. Comparing theory and experiment we now have 
$a \approx 23 \, \mbox{\AA}$. 

We also note that
the charge order results in Fig 3 of the main text used $\lambda=1$ because a cluster Monte Carlo
algorithm was possible only for this value. However, we did find that the shape of charge order structure factor peak
was relatively insensitive to the values of $\lambda$ and $w$, while these parameters are more consequential for the diamagnetic susceptibility.


\begin{thebibliography}{}
\bibitem{neellro} In La$_2$CuO$_4$ there is long-range antiferromagnetic
order below 325K, but this is entirely due to the inter-layer exchange interactions.
\end{thebibliography}
\end{document}